# SharkGraph: A Time Series Distributed Graph System

Derong Tang, *Tencent*

**Abstract**

Current graph systems can easily process billions of data, however when increased to exceed hundred billions, the performance decreases dramatically, **t**ime **s**eries data always be very huge, consequently computation on time series graphs still remains challenging nowadays.

In current piece of work, we introduces SharkGraph, a (**d**istributed **f**ile **s**ystem) **DFS**-based time series graph system, used a novel storage structure (**T**ime Series **G**raph Data **F**ile) **TGF**, By reading file stream to iterate graph computation, SharkGraph is able to execute batch graph query, simulation, data mining, or clustering algorithm on exceed hundred billions edge size industry graph. Through well defined experiments that shows SharkGraph performs well on large-scale graph processing, also can support time traversal for graphs, and recover state at any position in the timeline.

By repeating experiments reported for existing distributed systems like GraphX, we demonstrate that SharkGraph can easily handle hundreds billions of data, rather than GraphX which met many problems such as memory issues and skewed distribution on graph traversal. Compared with other graph systems SharkGraph uses less memory and more efficiently to process the same graph.

## 1. Introduction

Graph processing and analyzing huge real-industry graphs is always one of the most difficult problems faced by engineers and researchers. For example, social networks, Web graphs and transaction graphs are particularly challenging to handle, because these graphs are very huge and cannot be easily decomposed into small parts. When processing them we are facing skewed distribution problems like big nodes in social networks.

Most current graph systems only support one version in edge and vertex for graph mining, clustering and so on, but hardly simulate graphs at any position in the timeline. Such scenarios in simulation of graphs provoked us to ask a question: Would it be possible to combine **time series + graphs**? Based on the time series we can simulate an arbitrary state of the graph. Time series graphs will bring a lot of problems like storage capacity, even more serious skewed data distribution, more complex graph iteration and aggravate distribution problems. Multi version of graphs could increase the storage capacity dramatically, Time series graph storage systems will be more challenging. Unfortunately, when the graph storage capacity becomes bigger and bigger, graph computation will be less efficient.

Time series graphs are a pretty new field in graph computing, and we are facing a more serious above-mentioned problem, because time series graphs store multi-version data of graphs, we need to get the graph state at any position in the timeline. More generally, we also need to consider the problems of distributed systems like fault tolerance, how to manage a cluster robust and often unpredictable performance by various workloads in a machine. Time series graph stores multi-version of vertex and edge, edge struct in a time series graph simply like src, dst, edge_type, timestamp, attributes, and vertex also has a snapshot on each update, each node state struct just like vertex_id-> array(timestamp, attributes), we should record each update of graph in timeline, thus can easily get every snapshot of time series graph.

Sharkgraph is designed for time series graph data, it can simulate a whole graph state at any position in the timeline. Firstly, SharkGraph storage is based on **DFS. We** present a novel graph file structure for time series graphs, our compute iteration of graphs based on edge traversal which is accelerated by sort file stream. It uses less memory, becomes more robust, predictable performance, and also stays out of memory issues. Secondly, the computation model of **SharkGraph**, state-of-the-art, which is vertex-centric mode, most notably of them is **Pregel [1]** and **GraphLab [2]**, the difference between them, one is synchronous way and the other is asynchronous, SharkGraph choose **GAS(G**ather, **A**pply, **S**catter) **[3]** , by the consideration of more simple to implement, fault tolerance support and high parallelism.

The outline of our paper show as follows, We introduce the graph file structure in Section 2, then describe the how to achieve better file compression in Section 3, then SharkGraph computational model and challenges we met listed in Section 4. We evaluate SharkGraph on very large graphs that exceed a hundred billion edge size, using a set of algorithms to evaluate its performance such as graph cluster, graph mining, graph query and machine learning.

Our contributions:
- **T**ime Series **G**raph Data **F**ile which shortens as **TGF**, a novel graph file structure on a distributed file system, it could support edge and vertex for time traversal in large scale.
- Graph computation model on file stream, edge traversal of SharkGraph is based on distributed edge file stream in sort mode, graph computation depends on TGF. SharkGraph has the ability to efficiently process time series graph data in batch.
- System design, and implementation of SharkGraph. We use the **HDFS** as a base file system to implement **TGF**.

## 2. Graph File Structure

In this section, we start by describing the edge and vertex file structure, then present the partition strategy in SharkGraph, and continue by arguing why this structure is efficient to perform computation on graphs.

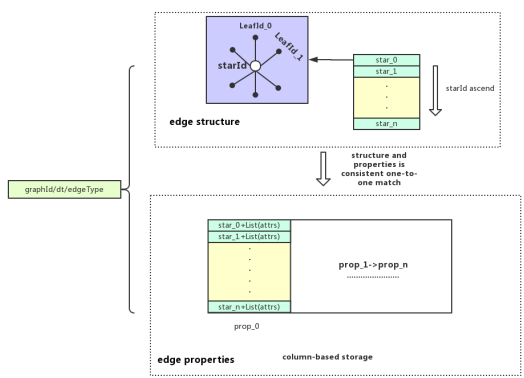

Figure 1. The edge file structure in SharkGraph

### 2.1. Edge

Edges in SharkGraph is a tuple object of five elements, each is src_id, dest_id, edge_type, attributes, we split every edge into two parts, first is the structure data contains src_id, dest_id, another is the attribute data contains timestamp, user defined attributes.

**Star Edge Structure:** There are many duplicate source ids in our edge structure because the real-industry graph is quite centric, skewed in a few big nodes. Thus SharkGraph grouped edges into star-like structures such as one source id to multi destination ids. Star structure is the minimum storage unit in edge structure, we designed this for the purpose of building graph index easily, and also compression friendly. We will introduce the design of the block in the next section.

**Partition strategy:** In SharkGraph uses source, destination, timestamp as a factor. Prior to presenting the reason for this strategy, we discuss some alternative strategies and the relevant problems they will encounter. If we use a single element of source or destination as partition key, the edges containing the same src or dst will go to the same partition, it will intensify the skewed distribution problem when met with big nodes. But its advantage is that we can simply route to the edge partition by just vertex id. The alternative partition strategy is 2-dimension, which uses source id and destination id, it could effectively scatter big nodes into different parts, however 2d introduced a problem of how to traversal from vertex partition to edge partition, thus we need to build index files for routing. This strategy can solve most problems of the usual graph. The circumstance of real-industry graphs typically have extremely super centric problems, such as a social network, where some node would like to communicate with the same person very frequently, consequently producing much more edge versions with the same source id and destination id. In normal graph process situations, we just need to record one version of the edge with the same src and dst, however in time series graphs we must record all of them, as a result we should provide the snapshot of the graph at any position of timeline. Based on this consideration, we introduce the 3-dimension strategy which uses source id, destination id and timestamp. We split the timestamp into hours, and partition data by source id, destination id and hour, which could distribute the data more evenly. As real-industry graphs typically have a very skewed distribution, we use this strategy to solve data distribution problems effectively. It stores big nodes more evenly, and also distributes the computation on it, significantly improving the efficiency of graph computation.

**Edge attributes**: We would like to store attributes in a graph, because a time series graph is used to aggregate in some property dimension. For example, we'd like to retrieve people who were above age 16, and how many friends in his social network. This is a typical scene in graph query. Finally we build the graph edge containing five parts including source, destination, edge type, attributes. We use a column-based model to store the attributes for better compression and column pruning support in query. Column-based attributes would make sense to compress in block, because our data in the same column has the similar data type. As a result we could do some special compression. For instance the timestamp field has a very efficient method to compress, the timestamp is a long type of data with 8 bytes. But in our graph the timestamp is always in ascending order. So we could only store the offset between 2 timestamp to save bytes in store which we called offset compression which will easily save half of space. We will present our compression algorithm in detail at section 3. When we query the attribute of edges, it possibly only need one of them, column-based attribute would significantly improve the read data scale by pruning, because it just need one type files. Edge attribute file structure one-to-one matches the star structure. We arrange the leaf edge attributes in the same order. We now support several data types including int, double, string, long for any defined attributes, and use specified type compression for all of them. A high-level illustration of file structure is given in Figure 1.

**File organization:** When we read the time series graph in a specified period, The related edges in this period would be read from dfs, filtered by timestamp in edge structure. In order to read edge files efficiently, we set some predefined partition fields such as graph id, date, edge type like **HIVE Format [4]**. Finally SharkGraph file path looks like this dfs://graphId/dt/edgeType/*** shown in Figure 3, we can easily find the needed edge collections in a time period by file path. Simultaneously it can filter out useless edge files according to date and edge type. The file header of the edge file contains the index based on the range of star structure id, the detailed introduction of the index will appear in section 2.4. Based on file path and index filtering, we reduce the amount of IO, so it can process data more rapidly in each graph iteration.

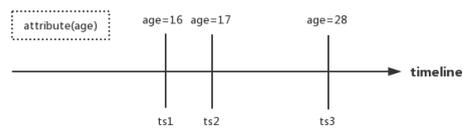

Figure 2. vertex attribute update timeline

**Global To Local:** Graph vertex id is a long type of data that has 8Byte, because we want to support hundreds of billions of data, so we arrange the storage space very carefully. Furthermore, the time series graph is always larger than usual, the gains will be very impressive. Prior to presenting the global to local structure, we discuss why should this work effective, graph data is split in many partitions, vertex id is a unique identifier for large scale data, so it use 8 byte to identify them, but in one partition it can hardly exceed the range of 4 byte which is approximately 2 billion, in the same partition we just need 4 byte to handle it, this also saves half storage space, on the other hand we need a mapping between local vertex id to global vertex id, global to local is proposed based on this situation, next we will discuss why it's advance by using this method, if the same vertex id appears several times in graph edges, that could save more storage. In time series data, it contains much more duplicated vertex ids in edge structure, measured in real-industry data, this would save about 20%-30% space in production. Therefore we introduce a novel mapping structure from global id to local id, global id is the origin 8 bytes id, and local id 4 bytes of local partition. We will store extra space for global id in the file system.

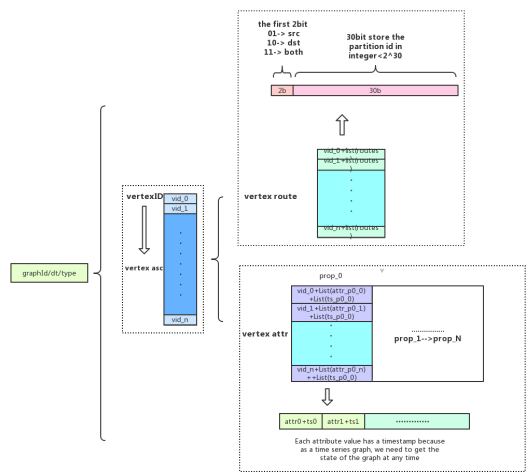

Figure 3. overview of vertex file structure

## 2.2. Vertex

Vertex file contains route and attributes, route file used to link to the matched edge partition, our implementation SharkGraph which traversal vertex along direction edge through the route table, furthermore vertex attributes is also a multi-version data.

**Vertex route:** The implementation of SharkGraph computation based on vertex-centric model, our query or algorithm start from vertex, so the critical problem is how to index vertex and edge with best efficiency for graph traversal, we proposed a novel route file structure, this vertex route generated by edges from the same time period, one edge contains two vertex ids, route info contains two parts, one is the location info **SRC, DST, BOTH** in the edge, next is the edge partition id. For the purpose of reducing storage space, we use 4 byte integers to store two parts info. The first two bits that 01 means SRC, 10 means DST, and 11 means BOTH appearance. The other 30 bits is representation of partition id, therefore our graph partition size must be less than 2^30, it totally fulfills the need. The flow of graph traversal like this, we read the vertex file first, then pop up route info contains the partition id, next shuffle the vertex to the edge partitions, and then retrieve the specified partition edges and filter by vertex id, finally we get the destination side of the edges. Graph iteration repeats the above process in order to traverse the graph. In the route file, we only store the route id, bring up the vertex id, and retrieve vertex attributes by the same vertex id, which could reduce the cost of storage extremely.

**Vertex Id**: In this file structure, we only have the long type id, for the purpose of index route and attribute file, the id in the sequence is distinct from others, so the global to local algorithm could not optimize the storage efficiency. A major advantage of sequence data is that we could use a more efficient compress method like **DFCM [5]**, which was widely used in many state-of-the-art databases. Vertex id in file structure is assigned ascending order, adjacent numbers have more similar bits, so our compression algorithm can achieve better efficiency. The route and attribute file have the same order as the id file.

**Vertex attributes:** In time series graph, vertex attribute may be updated at every moment, we must snapshot all of them for the purpose of time traversal, each time state of vertex attribute, we also use column-based mode, each type of attribute corresponds to a timestamp, this is used to mark when the property is updated, therefore we can trace back the vertex attribute state at any time. We now describe a simple example to explain how this file works, based on Figure 2, we have the attribute named age contains three version of values in three moment.The storage structure is composed of two sequences age [16,17,28] and timestamp [ts1,ts2,ts3]. This enables a more compact data structure for storage. When our graph time state is between ts2 and ts3, by this time point, we see the vertex attribute in ts2, This makes graph vertex trace back available.

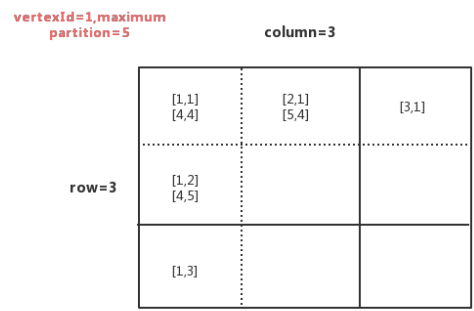

Figure 4. 2-dimension partition method

### 2.3 Graph Partition Strategy

A type Graph structure contains two parts, vertex and edge ,the partition strategy of graph also has two parts, vertex partition strategy and edge partition strategy.

We use vertex id as the partition key in vertex strategy, vertex partition can be determined only by vertex id, this strategy is designed for finding the partition of vertex quickly.

Prior to the current edge partition strategy, we want to introduce two questions of the graph-cut problem. In real-industry graphs often have some skewed vertex which have an excessive number of edges in graphs like social relations. Many famous people have much more followers than long tails. So our partition strategy should take the compute distribution into consideration, if most nodes are gathered in one partition, the computation of the graph appears to be a long-tail problem. The solution is to scatter the big nodes in many partitions, but it will also lead to more cross-partition communication, decreasing the performance of graph iteration. Therefore choosing a partition strategy requires balance the impact of distribution and communication overhead. In time series graphs, skewed distribution is more serious. So in SharkGraph we choose src, dst, timestamp , and three dimensions as the partition key to scatter edges into different partitions; we don't want to see the edges with the same source scattered over too many partitions. We use a novelty algorithm to guarantee edges are scattered in limited partitions. Firstly we have n^2 partition, organize partitions into a matrix of n*n, secondly we use src to obtain the row id, and destination id, timestamp to obtain the column id. In the worst case, it will only be scattered in 2n-1 partitions. A high level illustration of partition strategy is shown in Figure 4.

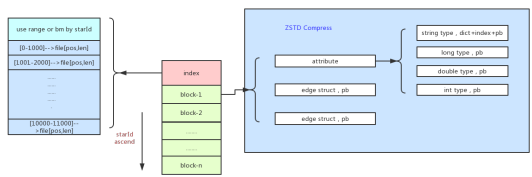

Figure 5. Index and compression in edge

## 3. Graph File Compression And Index

### 3.1 Graph Index

In file systems, in order to read files efficiently, we build indexes for graphs. We can easily find the right position of the vertex or edge. There are two ways of indexing, range-based index or bloom-based index. We'll illustrate this in detail.

**Range-based index:** We index our vertex or edge using the id column, because it is a long type struct and easy to sort. In a vertex-liked file (route, attribute), we use vertex id as the key, and put the index in the file header. Vertex file in SharkGraph is made up of file blocks followed one by one. The order of bytes in the file is in vertex id ascending order, so every block contains a vertex id range. When the input is a set of vertex id, first of all, retrieve the index info into memory, and find the block where actual data is located, and then skip to the right byte position, read file block. Our file blocks are ordered to benefit from sequential reading in disks, improving disk IO performance.

**Bloom index**: Bloom filter is a space-efficient probabilistic data structure, which is well known by low cost and efficient filtering methods. So another kind of indexing is based on bloom filters. Firstly we add all the vertex id of the block into a bloom filter, then transform it to bytes, our index is constructed from all these bloom filters. The index way of the file is basically the same as above, but we use bloom filter to confirm if the vertex id is in this block, A high level illustration of graph edge index is given in Figure 5.

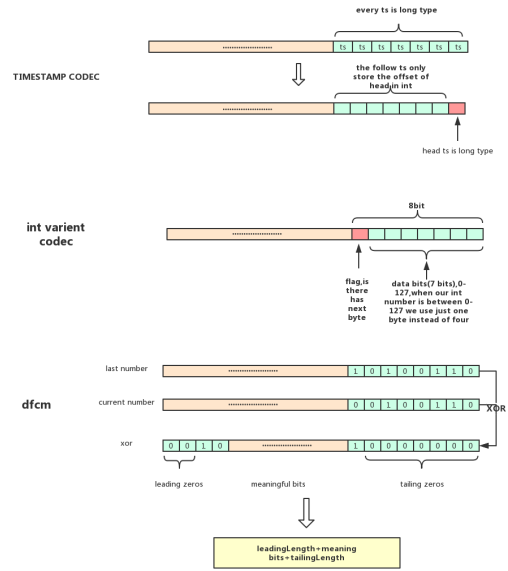

Figure 6. The variant compression in SharkGraph

### 3.2 Graph Compression

The file storage in time series graphs is always at large scale, we have to consider saving storage costs, so compression of graphs is seriously important. To enhance the benefits of compression ,our minimal unit of compression is graph file block.

**Edge Compression:** star block in edges and vertex id block in vertices. Like above figure 5, in edge files, many star structures constructed in the file block, our edge structure in star-graph just like this star id->sequence of leaf id, in each partition, firstly we use a global to local algorithm to reduce the id from 8 bytes to 4 bytes. Therefore we could use the variant codec to reduce the int series size. Then, based on some general compression algorithms, applied to the edge block file. In star graph attribute files, which are column-based, each file is one column of properties, supporting four basic data types long, int, string, double. The benefit of column-based file is compression friendly and column pruning in the graph process. On compression we could use specified serialization method to each type of properties before general compression algorithms. We use DFCM [5] long/double series compression method to long/double type attributes, variant codec to int type attributes, and dictionary method for string values, just shown In Figure 6, We briefly introduce two compression methods. After the above pre-process, we will do some general compression based on edge blocks such as snappy or zlib.

**Vertex Compression:** Vertex files contain route and attribute, the structure of route file is array integers, introduced in the second segment. Also pre-process by variant codec. The attributes file is column based, the attribute in the time series contains two parts. First is the values and the other is timestamps. In timestamp series we must use 8 bits to mark the time point, but in one block we just need to save the first timestamp in long, else only need to use 4 bits to store offset between first timestamp. After the above process we also use a general compression algorithm to compress the vertex block.

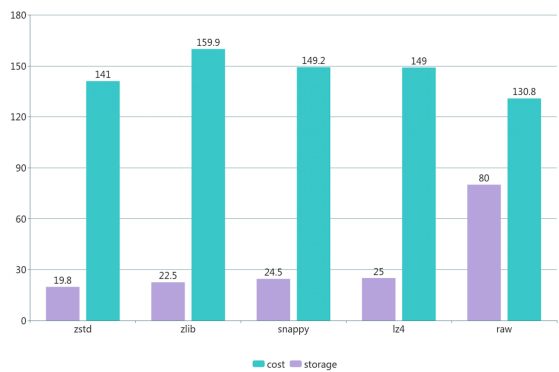

Figure 7. The compression cost and reduce space in different algorithm

**Compression Evaluation:** We compare five types of compression algorithm in the same real-industry data with metrics from time cost and space usage. Obviously, the zstd algorithm from facebook can achieve better results in the trade-off of time and space, which is shown in figure 7.

## 4. Computation on SharkGraph

### 4.1 vertex centric computation model

Vertex centric computation on graphs needs to be thought like a vertex. Pregel is well known as a vertex-centric Programming Model. The high-level organization of Pregel programs is inspired by Valiant's Bulk Synchronous Parallel model [6]. Pregel computations consist of a sequence of iterations, called supersteps. During a superstep the framework invokes a user-defined function for each vertex, obviously in parallel. The function specifies behavior at a single vertex **V** and a single superstep **S**. It can read messages sent to **V** in superstep **S** − 1, send messages to other vertices that will be received at superstep S + 1, and then modify the state of **V** and its outgoing edges. Messages are typically sent along outgoing edges, but a message may be sent to any vertex whose identifier is known.

```
1 Graph Tranversals
INPUT: input_vertex
OUTPUT: output_vertex
 1:  for pid = 1 to partitionSize do
 2:      vertex_memory ← LoadVertex2MemByPid(pid)
 3:  end for
 4:  Current_vid ← input_vertex
 5:  loop(Tranversal_once ← Tranversals)
 6:      for vertex ∈ Current_vid do
 7:          // get the route pid to edge partition
 8:          Current_route ← GetRouteFromVertex(vertex_memory)
 9:      end for
10:      // shuffle the input vertexes by route pid
11:      ShuffleVertexByRoutePid(Current_vid, Current_route)
12:      for pid = 1 to partitionSize do
13:          EdgeFileStream ← LoadEdgesFromDistributedFs(pid, CurrentPartition_Vid)
14:          filter EdgeFileStream by Current_vid
15:          dst_vertex ← EdgeFileStream
16:      end for
17:      Current_vid ← dst_vertex
18:  end loop
19:  return Current_vid
```

Algorithm 1. the traversal algorithm in SharkGraph

Our SharkGraph also uses a vertex-centric computation model, and supports the **Pregel** API, the implements support graph traversal is based on sorted file stream. The details of the execution process is in Algorithm 1. The input of iteration is a set of vertex ids, the output of traversal is the destination vertex id of edges, in every super step we loop this, firstly retrieve route info from input vertex ids, then shuffle all of the inputs by route pid to the right edge partition, and then we read all the edge file contains struct and selected attributes, search in edge index, then skip to the location of right block, in block processing, we filter the edge with input source ids, after all we could get to the destination in current step.

### 4.2 Update of vertices

It is quite common that the amount of vertices in a graph is relatively small compared to the amount of edges, so there is sufficient memory to store the array of vertex values. In most cases, we just need update vertex values in memory, so at the start of progress we load the vertices data into memory, then repartition it by vertex id, at each iteration, the input vertex ids, find the vertex route in memory, then read the edge in above step, the destination id receive all the message and apply the transform, the values in memory updated, waiting for next super

step(volt of halt). Next section we would illustrate all the update methods in PageRank and SSSP.

## 5. Conclusion

SharkGraph is a leading solution for time series graph, it use advanced data structure to improve the store performance and reduce the space, with specified compression method, we save 30% space and improve the batch traversal performance to 20% on real-industry graph, also depends on the **GAS** computation model it can solve different graph compute problem, like traversal, pagerank, sssp. Compared to GraphX **[3],** with less memory usage, and improved 3-degree query performance about 3 times in highly skewed distributed data.

## Reference

[1] Malewicz G, Austern M H, Bik A J C, et al. Pregel: a system for large-scale graph processing[C]//Proceedings of the 2010 ACM SIGMOD International Conference on Management of data. 2010: 135-146.

[2] Low Y, Gonzalez J E, Kyrola A, et al. Graphlab: A new framework for parallel machine learning[J]. arXiv preprint arXiv:1408.2041, 2014.

[3] Xin R S, Gonzalez J E, Franklin M J, et al. Graphx: A resilient distributed graph system on spark[C]//First international workshop on graph data management experiences and systems. 2013: 1-6.

[4] Thusoo A, Sarma J S, Jain N, et al. Hive: a warehousing solution over a map-reduce framework[J]. Proceedings of the VLDB Endowment, 2009, 2(2): 1626-1629.

[5] Burtscher M, Ratanaworabhan P. High throughput compression of double-precision floating-point data[C]//2007 Data Compression Conference (DCC'07). IEEE, 2007: 293-302.

[6] Gerbessiotis A V, Valiant L G. Direct bulk-synchronous parallel algorithms[J]. Journal of parallel and distributed computing, 1994, 22(2): 251-267.

[7] Page L, Brin S, Motwani R, et al. The pagerank citation ranking: Bring order to the web[R]. Technical report, stanford University, 1998.